\documentclass{optica-article}

\journal{opticajournal} 

\articletype{Research Article}

\usepackage{lineno}

\begin{document}

\title{$10^{-21}$-Level Instability Frequency Dissemination over 2,067 km noisy Telecommunication Infrastructure}

\author{Fa-Xi Chen,\authormark{1,2$\dagger$} Li-Bo Li,\authormark{2$\dagger$} Jiu-Peng Chen,\authormark{1,2$\dagger$} Kan Zhao,\authormark{2} Jian-Yu Guan,\authormark{1} Yang Xu,\authormark{1,3} Lei Hou,\authormark{1} Fei Zhou,\authormark{1,2}  Cheng-Zhi Peng,\authormark{1,3} Qiang Zhang,\authormark{1,2,3*} Hai-Feng Jiang,\authormark{1,3*} and Jian-Wei Pan\authormark{1,3*}}

\address{\authormark{1}Hefei National Laboratory, University of Science and Technology of China, Hefei 230088, China\\
\authormark{2}Jinan Institute of Quantum Technology and CAS Center for Excellence in Quantum Information and Quantum Physics, University of Science and Technology of China, Jinan 250101, China\\
\authormark{3}Hefei National Research Center for Physical Sciences at the Microscale and School of Physical Sciences, University of Science and Technology of China, Hefei 230026, China}



\begin{abstract*} 
The realization of ultra-stable optical frequency transmission through fiber networks is critical for advancing global optical frequency standards and enabling applications such as redefining the second in the International System of Units (SI), geophysical sensing, quantum network construction, and fundamental physics experiments. However, achieving high-reliability and low-instability optical frequency carrier transmission links over distances exceeding thousands of kilometers remains technically challenging, thereby limiting the scalability and reliability of such networks. In this study, we experimentally demonstrate that the noise accumulation in long-distance optical links can be mitigated by narrowband purification of the optical signal’s phase noise, enabling optical links of theoretically unlimited length. Additionally, we implemented digital optical phase measurement and feedback technology to calibrate noise compensation deviations caused by inconsistencies in round-trip optical frequencies, enhancing link stability. By adopting digital phase measurement instead of traditional phase detectors, we expanded the dynamic noise tolerance range of the optical phase-locked loop, significantly improving system reliability. Ultimately, on a 2,067 km telecommunications fiber link with a noise level exceeding 5000 $rad^{2} Hz^{-1} km^{-1}$, we achieved an optical frequency transfer with a daily instability of $2.9 \times 10^{-21}$ without experiencing any optical cycle slips maintaining continuous operation for four days. This work establishes a technical foundation for leveraging existing fiber resources to construct global-scale optical frequency standard networks.


\end{abstract*}

\section{Introduction}
 Optical atomic clocks, heralded as the pinnacle of precision metrology, have achieved unprecedented uncertainties of $10^{-19}$~\cite{aeppli2024clock}, surpassing the performance of microwave-based cesium standards by over two orders of magnitude~\cite{zhiqiang2023176lu+,brewer2019al+,aeppli2024clock,dimarcq2024roadmap}. This leap in precision has positioned optical clocks at the forefront of redefining the International System of Units (SI) second~\cite{dimarcq2024roadmap,lindvall2025coordinated}, a milestone anticipated within the next decade. Yet, this transformative leap—which promises to revolutionize fundamental physics, quantum networks, gravitational wave astronomy and geodesy~\cite{yu2020entanglement,chen2020sending,pompili2021realization,clivati2022coherent,delva2017test,roberts2020search,collaboration2021frequency,dix2023experimental}—confronts a critical barrier: the stark disparity between the extraordinary stability of optical clocks and the inadequacy of global time-frequency (TF) transmission infrastructure. To harness this metrological leap, next-generation TF networks must not only match optical-clock stability but also maintain robustness across global scales, operating seamlessly amid the environmental noise, latency, and infrastructure variability inherent to real-world deployment.


 The future TF infrastructure is envisioned as a hybrid system integrating satellite-based free-space links for intercontinental backbone connectivity with terrestrial fiber-optic networks spanning thousands of kilometers~\cite{shen2022free,caldwell2023quantum,caldwell2025high}. While satellite systems demonstrate resilience to extreme losses (89–100 dB) and femtosecond-level instability~\cite{shen2022free,caldwell2023quantum}, fiber optics remain indispensable due to their universality and unmatched precision. Europe leads in long-haul fiber-TF technologies, leveraging low-noise, buried, dedicated fibers to achieve instabilities below $10^{-18}$ over distances up to approximately 2,000 km~\cite{droste2013optical,calonico2014high,chiodo2015cascaded,koke2019combining,schioppo2022comparing}. Innovations such as Brillouin amplification (Physikalisch-Technische Bundesanstalt, PTB)~\cite{terra2010brillouin,koke2019combining}, cascaded noise compensation (Systèmes de Référence Temps-Espace, SYRTE)~\cite{fujieda2009coherent,lopez2010cascaded,chiodo2015cascaded}, and noise-purified filtering (Istituto Nazionale di Ricerca Metrologica, INRiM)~\cite{calonico2014high,schioppo2022comparing} enhance performance. However, reliance on dedicated or low-noise fibers (typically <~1 $rad^{2} Hz^{-1} km^{-1}$ at 1 Hz) limits global scalability. Notably, only one transmission—PTB’s 1,400 km dedicated low-noise system—has achieved uninterrupted operation for more than a few days~\cite{koke2019combining}, underscoring the urgency of advancing robustness in noisy, non-dedicated fiber environments.
 
 In contrast, rapidly urbanizing regions like East Asia, where many fiber segments are subjected to amplified mechanical and thermal perturbations, exhibit dynamic noise levels ranging from tens to hundreds of $rad^{2} Hz^{-1} km^{-1}$ at 1 Hz or higher~\cite{akatsuka2020optical,deng2024coherent,hu2021performance}. Such noise overwhelms conventional delayed optical phase-locked loops (OPLLs), whose limited dynamic noise rejection capabilities lead to frequent link interruptions, relegating TF transmission to short-duration frequency comparisons~\cite{akatsuka2020optical,deng2024coherent,hu2021performance}. To address the stability, reliability, and scalability challenges of fiber-based TF links, we conducted work in two key areas: (1) Adopting purification and filtering techniques~\cite{calonico2014high,schioppo2022comparing} to suppress phase noise accumulated during long-distance transmission enables optical fiber links to be extended to greater distances. Dividing a long link into cascaded shorter segments (typically 200–300 km apart) using relay stations enhances noise suppression feedback bandwidth (to several hundred hertz). However, as distance increases, noise accumulates exponentially near the feedback bandwidth frequency (tens to hundreds of hertz) for each segment, leading to frequent out-of-phase-lock events that limit reliability. To overcome this, we employed a loosely phase-locked ultra-stable laser to purify phase noise near the feedback bandwidth, enabling link extension without compromising reliability. Experimental results show that performing purification every 1,000 km keeps accumulated noise within the system’s safe operational threshold, even in high-noise fiber networks. (2) Leveraging digital circuits to record optical phase changes caused by link noise and perform real-time compensation enhances system stability and reliability. First, this approach records the actual compensation amount, enabling correction of small deviations in link noise compensation due to system phase asymmetry from round-trip optical frequency shifts, further improving stability. Experimentally, the peak-to-peak time deviation was reduced from 15 femtoseconds to below 6 femtoseconds after correction. We believe the system’s current long-term stability is primarily limited by fiber polarization mode dispersion. Second, the high-speed sampling of the digital phase recording and compensation method overcomes the dynamic range limitations of traditional phase-locked loop detectors, significantly improving system reliability.

\section{Long-distance TF distribution optical fiber link}

\begin{figure*}[ht!]
\centering\includegraphics[width=14cm]{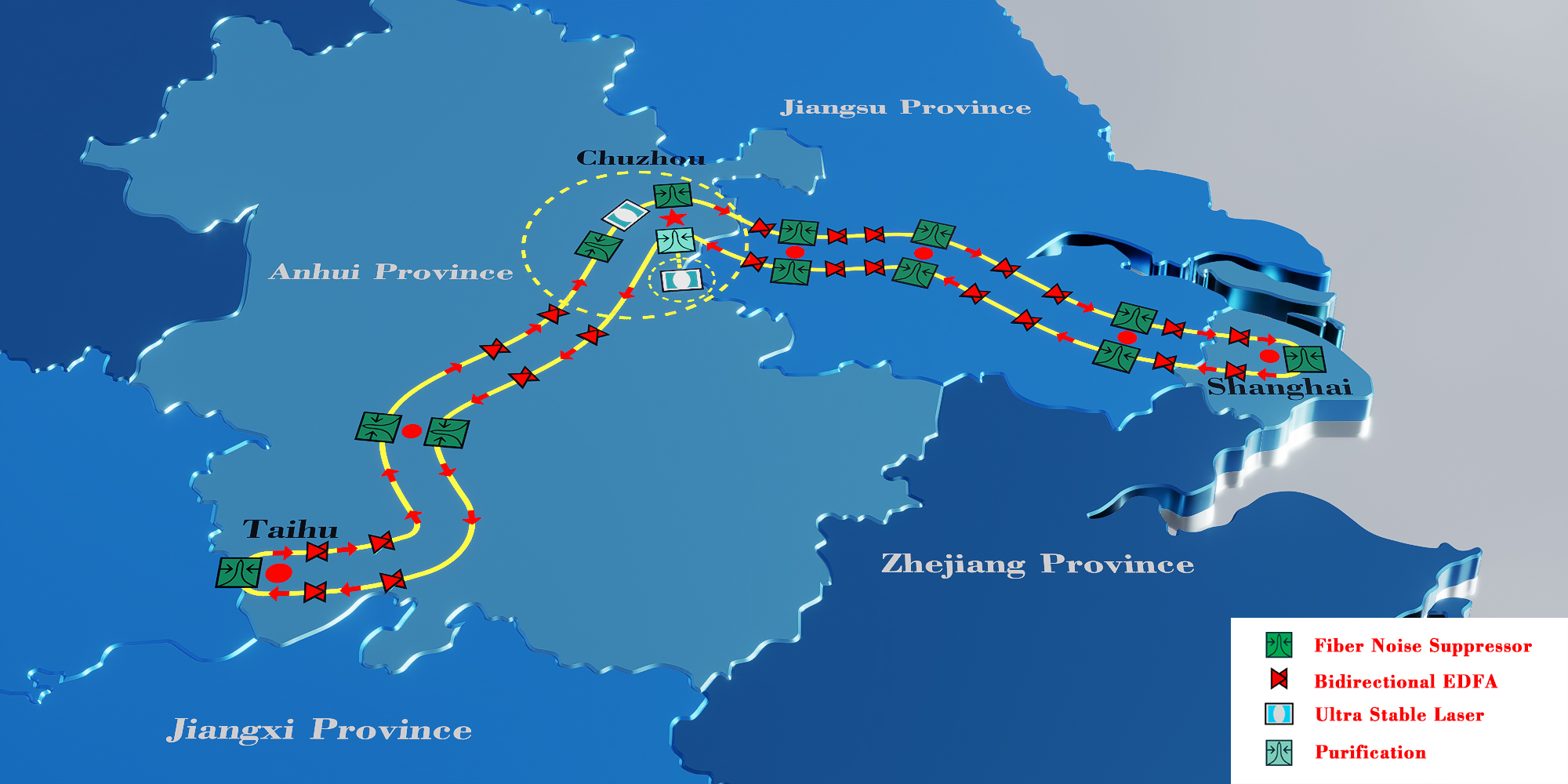}
\caption{Hierarchical framework for the 2,067 km TF dissemination. The long-haul TF dissemination comprises a transmitter and 12 hybrid repeaters—11 functional units and 1 core unit. The functional repeaters lock kHz-linewidth lasers, while the core unit locks an Hz-level ultra-stable laser. The entire optical link consists of a pair of 1,033.5-km-long telecommunications fiber networks within the same cable duct, traversing 18 metropolitan areas. The fiber pair forms a loopback formation by being connected at the station of Shanghai and Taihu. With the transmitter and all 12 hybrid repeaters operational, 13 cascaded noise suppressors are constructed to stabilize the full fiber span, enabling ultra-stable TF dissemination.
}
\label{Fig:Link}
\end{figure*}



The total length of the ring-type TF transmission link is 2,067 kilometers, consisting of two parallel communication optical fibers, each 1,033.5 kilometers long and housed within the same cable duct. This network connects 18 cities across China, with endpoints at the Shanghai station ($121^{\circ}33^{'}15^{"}$ E, $31^{\circ}8^{'}0^{"}$ N) and Taihu station ($116^{\circ}18^{'}7^{"}$ E, $30^{\circ}26^{'}21^{"}$ N). Dense Wavelength Division Multiplexing (DWDM) technology enables time-frequency signals to propagate within the ITU 34$\#$ channel, allowing the fiber to be shared for data transmission services in other channels, thereby reducing the cost associated with acquiring dedicated optical fibers. As depicted in Fig.~\ref{Fig:Link}, this optical fiber TF link employs a cascaded design, comprising one starting station and 12 relay stations (or repeaters): 11 standard functional units and 1 unit with phase noise purification capabilities. In Chuzhou, the starting station utilizes a Hertz-level ultra-stable laser (1550.12 nm) to generate both the signal light for the TF link and the local reference light required for stability analysis. The signal light travels eastward for 1,267 kilometers (via 7 functional repeaters) before returning to Chuzhou, where the purification repeater removes high-frequency noise from the optical frequency transfer signal. Subsequently, the purified signal propagates westward for the remaining 800 kilometers (through 5 functional repeaters), completing its return journey to Chuzhou.

The total link loss (including splices and connectors) is approximately 600 dB. To compensate for these link losses, 34 remotely controlled bidirectional erbium-doped fiber amplifiers (bidirectional EDFAs) were deployed, each providing a gain of 10 to 30 dB. For instance, the bidirectional EDFAs in Hefei used a 25 dB gain to counteract about 30 dB of attenuation. Each pump is flanked by optical filters with a bandwidth of 25 GHz to suppress out-of-band noise. 

On the urban-dominated routes, the phase noise at 1 Hz exceeds 5,000 $rad^{2} Hz^{-1} km^{-1}$, several orders of magnitude higher than the phase noise levels in previous long-haul TF systems ~\cite{predehl2012920,droste2013optical,calonico2014high,akatsuka2020optical}. Transient disturbances can cause phase shifts exceeding 10,000~$\pi$ radians within a 1 ms time interval.

\section{Repeater Architecture for Cascaded Noise Suppression in TF Dissemination}

Each repeater comprises two functional units: a reception unit and a transmission unit. The reception unit reflects a portion of the incident light for noise compensation in the upstream link while simultaneously receiving and purifying the laser signal transmitted from the preceding stage. The transmission unit employs the regenerated local laser signal as its source, utilizing a Michelson interferometer structure to measure and compensate for fiber-induced noise.

\begin{figure}[ht!]
\centering\includegraphics[width=14 cm]{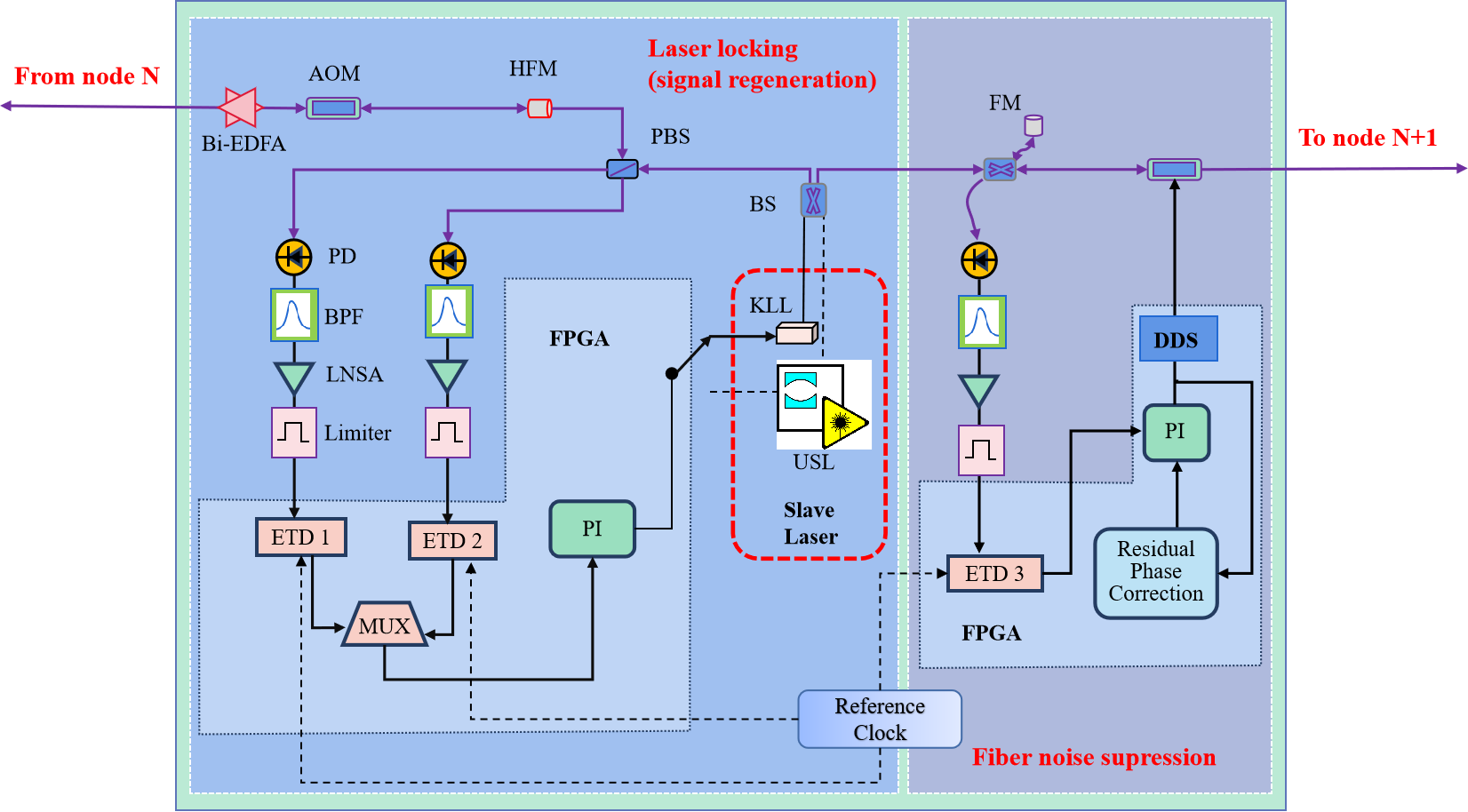}
 \caption{Hybrid Analog-Digital OPLL Architecture for Noise Cancellation and Laser Locking. All OPLLs employ an electronically controlled approach using a hybrid analog-digital control architecture. The analog processing module converts the heterodyned sinusoidal signal into a square waveform with stabilized amplitude, ensuring robust signal conditioning, while the digital processing module adaptively discriminates the phase noise of the conditioned signal and provides real-time feedback. HMF: Faraday half-pass plate. PBS: Polarizing Beam Splitter. BS: Beam Splitter. FM: Faraday mirror. PD: Photodetector.  MUX: Multiplexer. FPGA: Field programmable gate array. AOM: Acousto-optic modulator. USL: Ultra stable laser. KLL: kiloherts-linewidth laser. BPF: band pass filter. LNSA: low-noise self-adaptive-gain amplifier. PI: proportional-integral controller. DDS: direct digital synthesizer.
}
  \label{Fig:elec}
 \end{figure}

As illustrated in Fig.~\ref{Fig:elec}, the reception unit initially amplifies the input laser signal using a bidirectional erbium-doped fiber amplifier (bi-EDFA). Subsequently, the amplified signal undergoes frequency shifting via an acousto-optic modulator (AOM) operating at a fixed $-$40 MHz frequency. This frequency shift effectively isolates stray reflections within the fiber link between repeaters. The shifted signal is then directed into a half-transmitting and half-reflecting Faraday mirror (HFM). The transmitted beam serves as the reference signal, enabling the generation of a clean-spectrum and intensity-stable regenerated optical signal from the local laser through the use of an optical phase-locked loop (OPLL). In a standard repeater, the local laser source employs a kilohertz-linewidth laser, with the OPLL control bandwidth typically set to tens of kilohertz. In contrast, in a noise-purification repeater, the local laser source utilizes a hertz-level linewidth laser, and the OPLL control bandwidth is precisely set to approximately 3 hertz (see supplementary material for details).

The polarization state of the light in the communication fiber is susceptible to fluctuations caused by temperature gradients, vibrations, or fiber stress. These fluctuations can induce variations in the beat frequency signal between the received laser signal from the previous stage and the local laser, potentially leading to OPLL out of lock at the receiving end. To address this issue, a polarization beam splitter (PBS) heterodyne interferometer is employed to simultaneously measure the beat frequency signals between the two orthogonal polarization states of the received light and the local laser. This ensures that at least one beat frequency signal with a signal-to-noise ratio (SNR) exceeding the threshold required for normal OPLL operation can be generated Ref.~\cite{clivati2020robust}. A real-time algorithm dynamically selects the optimal signal for optical signal regeneration. 
In the transmission unit , the frequency shift used for noise compensation is 40 MHz, which is equal in magnitude but opposite in direction to the frequency shift applied in the reception unit. Consequently, the additional phases cancel each other out, reducing the accuracy requirement of the local reference source~\cite{lopez2010cascaded}.

All equipment has been designed for deployment in communication rooms and supports remote control via wireless communication networks. With the exception of the Chuzhou Central Station and the Shanghai/Taihu terminal stations, all other stations are equipped with two repeaters housed in a compact 19-inch rack-mounted chassis. Optical components are integrated into 1U units, while control electronics are accommodated in 3U chassis, ensuring seamless compatibility with standard infrastructure and long-term operational reliability. In these repeaters, fiber optic splices are restricted to a maximum length of 13 cm, and the mismatch in fiber lengths along the interference path is maintained within 1 cm to minimize the system's temperature coefficient. Furthermore, the optical components of the repeaters are centrally installed and actively temperature-controlled, achieving thermal stability with fluctuations below 10 millikelvins (standard deviation).

\section{Digitally Augmented Phase-Tracking System}

Robust phase tracking in TF dissemination systems requires unambiguous phase detection across all feedback loops, including laser locking in repeaters and inter-repeater fiber noise cancellation, as well as resilience to dynamic power fluctuations in beat notes. Conventional analog OPLLs, which rely entirely on analog circuitry, are inherently limited by their narrow phase-tracking dynamic range. These limitations compromise the stability and reliability of long-haul TF networks, particularly under real-world conditions involving noise and power variability. To overcome these constraints, we integrate digitally enhanced phase tracking into OPLL architectures. By combining adaptive digital signal processing with precision analog conditioning, this hybrid approach significantly expands the dynamic range (from tens to tens of thousands of radians within 1 $ms$) and enables real-time noise suppression, resulting in stable, robust, and scalable TF dissemination over extended distances—surpassing the performance limits of traditional analog systems.

\subsection{Hybrid Analog-Digital OPLL Architecture for Noise Cancellation and Laser Locking}

To optimize stability and noise immunity, all OPLL subsystems—including delayed OPLLs for sender/inter-repeater fiber noise suppression and OPLLs for laser locking in repeaters—employ hybrid analog-digital architectures. These architectures enable high-precision phase synchronization and broad dynamic-range operation.

As shown in Fig.~\ref{Fig:elec}, the sender/inter-repeater fiber noise cancellation loop operates as follows: During the analog signal conditioning phase, a photodetector (PD) converts the heterodyne optical signal into a sinusoidal electrical waveform. A 6 MHz narrowband band-pass filter (BPF) isolates frequency-correlated noise components. A low-noise self-adaptive-gain amplifier (LNSA) with a 60 dB dynamic range stabilizes amplitude fluctuations. A high-speed limiter then directly transforms the conditioned signal into a square wave, bypassing analog-to-digital converters (ADCs) to avoid quantization noise and amplitude-induced disturbances. In the digital processing phase, a precision event-timing module measures timing noise in the digitized waveform with <1 ns resolution, converting it into phase changes. A proportional-integral (PI) controller dynamically adjusts a direct digital synthesizer (DDS), which drives an acousto-optic modulator (AOM) for real-time phase compensation. The phase detection module and PI controller are integrated within a field-programmable gate array (FPGA), leveraging its reconfigurable logic for flexible processing.

The OPLL architecture is also applied to laser locking in repeaters by concurrently detecting two intensity-reciprocal heterodyne signals from a PBS using dual PDs. In the analog signal conditioning stage, each sinusoidal output is conditioned through a 4 MHz BPF, LNSA, and limiter chain to generate digital square waves. In the digital processing stage, event-timing-based phase discriminators extract phase differentials, with an intelligent algorithm selecting the signal channel exhibiting the superior signal-to-noise ratio (SNR). The selected phase error is processed by a PI controller, which adjusts the local laser’s frequency to achieve phase locking. This dual-control configuration ensures precise phase locking of the repeater station’s local oscillator to the incident optical signal, enabling robust regeneration fidelity even under dynamic environmental perturbations.

\subsection{Event-timing-based phase detection method}

Conventional analog phase discriminators in OPLLs, which compare reference and detected signal phases directly, are fundamentally constrained by a dynamic range of $0^{\circ}$–$360^{\circ}$ (one cycle).  Beyond this range, cycle ambiguities arise, degrading phase stabilization accuracy and rendering these methods unsuitable for multi-cycle, high-resolution tracking. Existing solutions—such as high-ratio frequency dividers in analog systems~\cite{predehl2012920,lopez2010cascaded,droste2013optical,chiodo2015cascaded} and phase unwrapping algorithms in digital systems~\cite{hu2021performance} —partially mitigate this issue but face critical limitations: Analog dividers require complex cascaded architectures to extend the dynamic range, introducing latency and noise. Digital unwrapping relies on phase continuity assumptions, failing under rapid phase variations. Both approaches struggle with monotonic dynamic ranges typically limited to below 1,000 $\pi$ radians. Such limitations are increasingly incompatible with the scalability and reliability demands of modern global infrastructure networks—particularly in environments with extreme or fluctuating environmental noise.

To overcome these limitations, we introduce a novel phase discrimination architecture that replaces conventional phase comparison with absolute event timing, leveraging time-to-digital converters (TDCs) for unbounded dynamic range and cycle-slip-free operation. The method operates as follows:

1. Timing Baseline: A reference clock establishes a global timing baseline synchronized to the rising edge of each cycle.

2. Signal Conditioning: The detected heterodyne signal is converted to a square wave, isolating timing information from amplitude noise.

3. Signal Timing: A TDC measures the absolute arrival time of the square-wave edge relative to the reference clock, resolving time differences with nanosecond  precision (equivalent to attosecond-level phase resolution at optical frequencies).

4. Phase Reconstruction: Time differences ($\Delta{t}$) are linearly mapped to phase deviations ($\Delta{\phi}$) using the relationship $\Delta{\phi}=360^{\circ} \times \frac{\Delta{t}}{T_{0}}$, where ${T_{0}}$ is the heterodyne signal period.

The system employs a 200 MHz reference clock to perform high-speed sampling of a 10 MHz signal, generating a 32-bit digital time code with a time-tracking dynamic range of $\pm1\times10^{10}$ nanoseconds (equivalent to a phase-tracking dynamic range of $\pm2\times10^{8}\pi$ radians at optical frequencies) and a single-measurement time precision of 5 nanoseconds (corresponding to 0.1 $\pi$-radian phase precision). This approach eliminates cycle ambiguities by decoupling phase tracking from the $0^{\circ}$–$360^{\circ}$ constraint, enabling multi-cycle stabilization across an effectively unbounded dynamic range. Unlike conventional methods, it remains robust to rapid phase variations and environmental noise, making it ideal for high-precision applications in unstable environments, such as urban infrastructure or long-haul optical networks.

\label{sec:Event-timing}

\subsection{Adaptive Phase Asymmetry Correction in Delayed OPLL Systems}

In an ideal symmetric delayed OPLL system, forward-path phase variations account for precisely half of the total round-trip phase fluctuation due to time-reversal symmetry~\cite{williams2008high,xu2021non}. However, the introduction of an acousto-optic modulator (AOM) at the receiving station breaks this symmetry by imposing a frequency shift ($\Delta\nu$) on the backward-propagating signal. This asymmetry creates mismatched carrier frequencies ($\nu_{f}\neq\nu_{b}$) between forward ($\nu_{f}$) and backward ($\nu_{b}$) paths, distorting the proportional relationship between forward-path and round-trip phase fluctuations. The resulting imbalance destabilizes conventional phase compensation schemes, which assume symmetric bidirectional propagation.

To address this, we embed a round-trip phase-compensation error-correction algorithm within the digitally augmented delayed OPLL for sender/inter-repeater fiber noise suppression that dynamically recalibrates the locking reference point in real time. Unlike fixed-reference systems, this architecture accounts for residual phase noise ($\Phi_{r}$) induced by asymmetric frequency propagation. The residual error is governed by: $\Phi_{r} =\delta\phi\times\frac{\Delta_{\nu}}{2\nu_{0}}$, where $\delta\phi$ is the incremental fiber Doppler phase noise, $\Delta_{\nu}=\nu_{f}-\nu_{b}$ is the bidirectional frequency offset, and $\nu_{0}$ is the nominal carrier frequency.

Accurate measurement of the phase difference between the controlled and reference signals is critical for sustaining long-term stability in phase-locked systems. Leveraging the FPGA’s flexible logic, the system performs real-time computation of the residual phase error $\Phi_{r}$ derived from cumulative round-trip fiber phase noise. By dynamically recalibrating the locking reference from a fixed zero point to $\Phi_{r}$, the architecture effectively decouples the effects of asymmetric propagation-induced bidirectional frequency shifts from the stabilization loop. This innovative approach significantly enhances long-haul TF dissemination, achieving unparalleled long-term stability.

\section{Experimental results and discussion}

To assess the precision of the long-haul TF dissemination, we conducted heterodyne detection by comparing the final received signal with the initial transmitted reference. The measurements were performed at full bandwidth to capture the overall noise contributions across the entire frequency spectrum.

\subsection{Phase noise analysis of the link}

\begin{figure}[ht!]
\centering\includegraphics[width=10 cm]{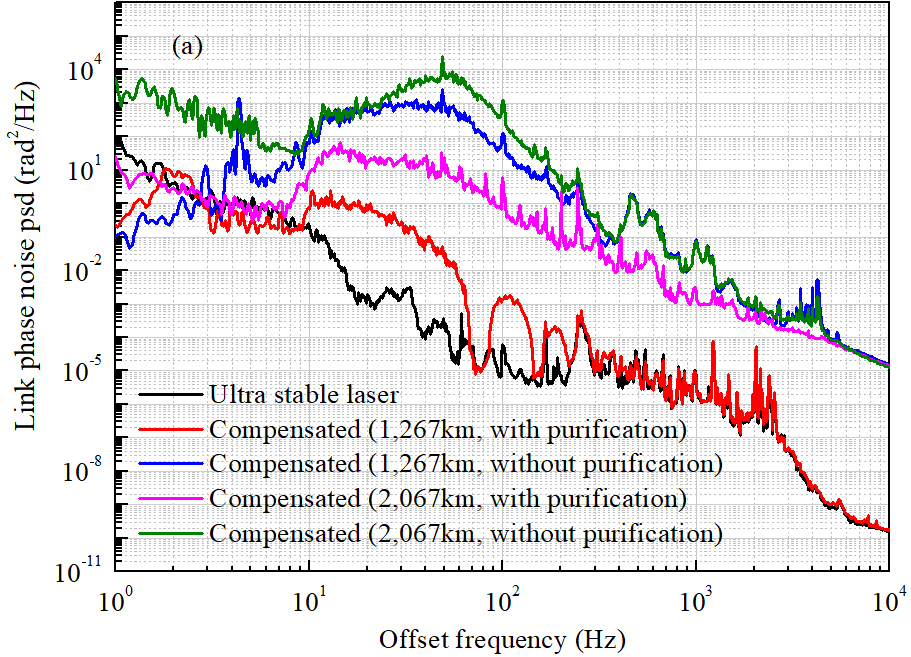}
\centering\includegraphics[width=10 cm]{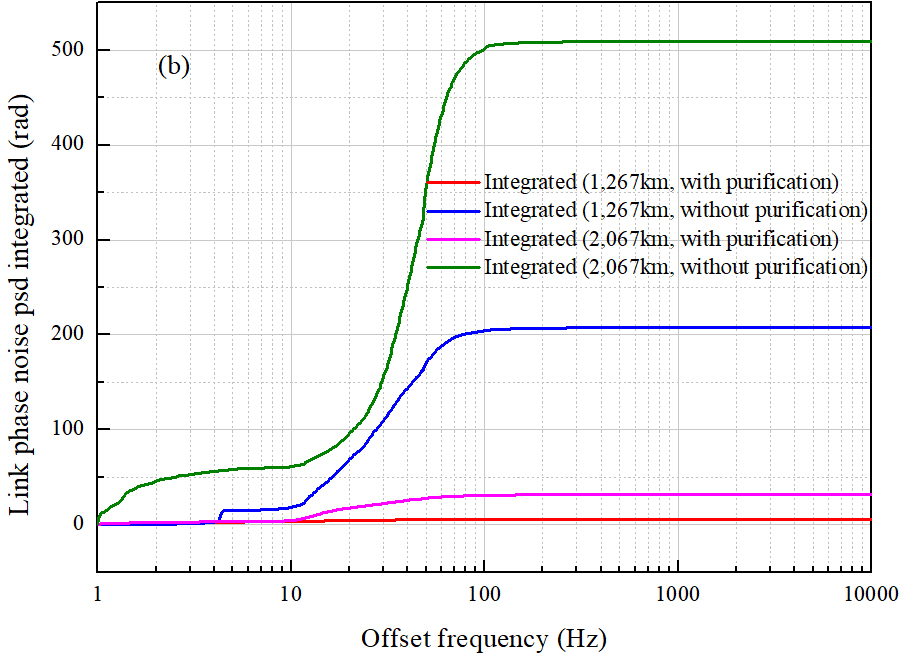}
 \caption{\textbf{(a)Phase noise power spectral density (PSD).} Black curve: Baseline phase noise between the transmitter source and purification laser. Blue curve: Compensated 1,267 km optical link (non-optimized) – phase noise PSD without purification, highlighting residual noise from cascaded repeaters. Red curve: Compensated 1,267 km optical link (optimized) – phase noise PSD with purification, demonstrating near-baseline noise levels through ultra-stable laser substitution. Green curve: Compensated 2,067 km optical link (non-optimized) – phase noise PSD without purification, showing amplified noise from servo-induced OPLL resonances. Pink curve: Compensated 2,067 km optical link (optimized) – phase noise PSD with purification, achieving near-complete suppression of cascaded OPLL noise from the preceding 1,267 km span. \textbf{(b) Integrated phase noise PSD.} Red/Blue curves: Compensated 1,267 km optical link with/without purification, demonstrating purification-driven noise reduction over intermediate distances. Green/Pink curves: Compensated 2,067 km optical link with/without purification, highlighting purification’s efficacy in suppressing cumulative noise across ultra-long spans. 
}
  \label{Fig:noise}
 \end{figure}

To assess the noise suppression efficacy of the digitally enhanced phase-tracking system, we evaluated phase instability in the frequency domain by computing the power spectral density (PSD) of phase fluctuations. The PSD was derived from the Fourier transform of the phase difference between the received signal and the transmitted reference signal. Figure~\ref{Fig:noise}(a) displays the phase noise PSD, while Figure~\ref{Fig:noise}(b) shows the corresponding integrated phase jitter. Activation of the cascaded fiber noise suppressors significantly enhanced stabilization at low frequencies. However, a key limitation emerged: servo-induced resonances in the cascaded optical phase-locked loops (OPLLs) amplified phase noise in the tens Hertz to a few hundrads Hertz range, compromising system stability even with all repeaters operational. This noise accumulates progressively, potentially exhibiting exponential growth over long distances, which undermines system reliability during extended transmissions.

For a 2,067 km link without a noise purification repeater, the additional phase jitter from 1 Hz to 10 kHz reached approximately 500 rad. To mitigate exponential noise growth in the tens to hundreds of hertz range, the control gain was deliberately reduced; however, the noise spectral density still peaked at approximately 10,000 rad²/Hz. At such elevated noise levels, optical phase locking suffered frequent phase slips due to instantaneous noise spikes and limited suppression capacity, with multiple occurrences daily.

To improve system reliability, we replaced the standard repeater at Chuzhou (1,267 km) with a noise purification repeater. This modification significantly reduced high-frequency noise, allowing an increase in control gain for more effective low-frequency noise suppression. Consequently, the overall additional noise level of the link improved substantially, reducing residual phase jitter from approximately 200 rad to $\thicksim$5 rad. The 2,067-km link with the purification station exhibited a phase jitter of $\thicksim$30 rad, markedly lower than the 200 rad observed in the 1,267-km link without purification. These results demonstrate that a single purification station can effectively mitigate noise accumulated over distances exceeding 1000 km. In principle, strategic deployment of purification stations could enable optical transmission links with virtually no distance limitations.

\subsection{The temporal delay variation of the optical link}

\begin{figure}[ht!]
\centering\includegraphics[width=10 cm]{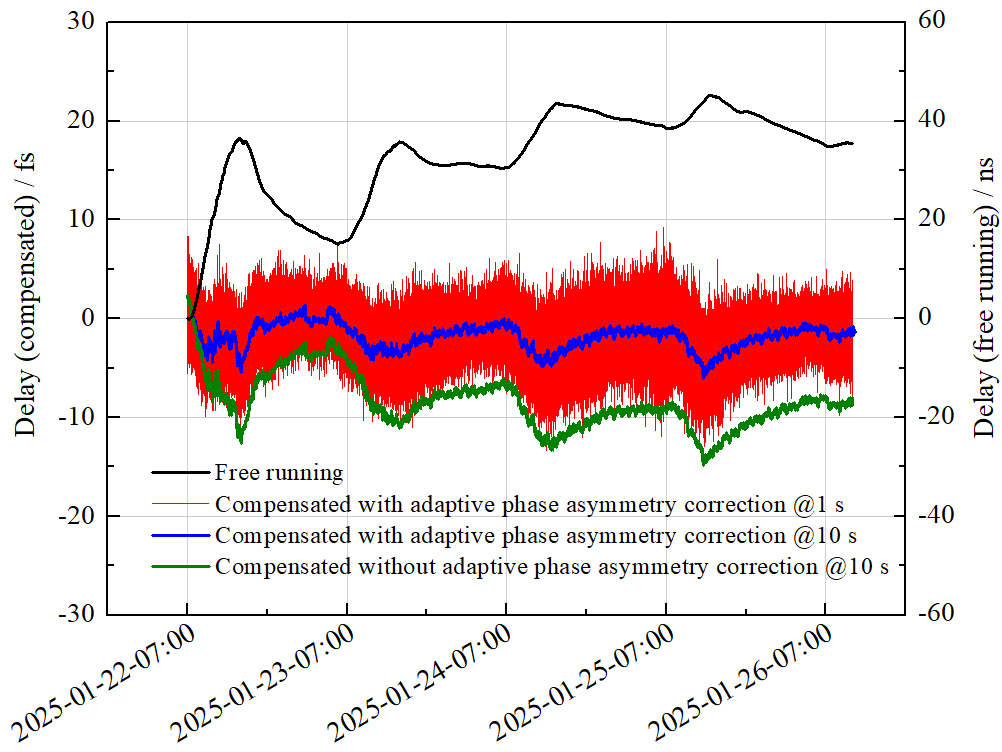}
 \caption{Temporal Evolution of Fiber Link Propagation Delay. Black Curve (Uncompensated/Free-Running): Baseline propagation delay with significant fluctuations due to uncontrolled thermo-mechanical noise in the fiber infrastructure. Green Curve (Compensated Without Adaptive Correction): Stabilized delay with residual drift caused by asymmetric bidirectional propagation-induced frequency shifts. Red/Blue Curve (Compensated With Adaptive Correction): Enhanced stability achieved by real-time adaptive phase asymmetry correction, minimizing residual drift. Both compensated states exhibit gradual drift driven by daily temperature fluctuations.
  }
  \label{Fig:delay}
 \end{figure}

Fig.~\ref{Fig:delay} illustrates the temporal evolution of phase transfer delay over a 4-day period for a 2,067-kilometer cascaded fiber link: (1) Free-running link (uncompensated, black curve): primarily driven by temperature-induced refractive index changes in quartz fiber, the peak-to-peak delay fluctuation reaches approximately 55 nanoseconds, correlating strongly with the 24-hour environmental temperature cycle; (2) Compensated link (without phase asymmetry correction, green curve): active stabilization suppresses symmetric noise components in the long-haul fiber, reducing delay variations to approximately 15 femtoseconds (peak-to-peak); (3) Compensated link (with phase asymmetry correction, red/blue curve): real-time correction of residual asymmetric errors induced by bidirectional propagation frequency shifts further reduces delay variations to approximately 6 femtoseconds (peak-to-peak), representing a 150\% improvement over the non-adaptive regime.

Although asymmetry-induced noise is effectively suppressed, residual drift persists, exceeding in-loop drift and strongly correlating with temperature fluctuations. This drift primarily stems from temperature-driven polarization mode dispersion (PMD) in the deployed fiber [30], which current compensation techniques cannot fully mitigate. Nonetheless, the optical link's instability meets the time-frequency (TF) transfer requirements for optical atomic clocks. For future high-performance TF links targeting sub-femtosecond accuracy, advanced methods to measure and compensate for PMD drift effects will likely be essential.

\subsection{Instability}

\begin{figure}[ht!]
\centering\includegraphics[width=10 cm]{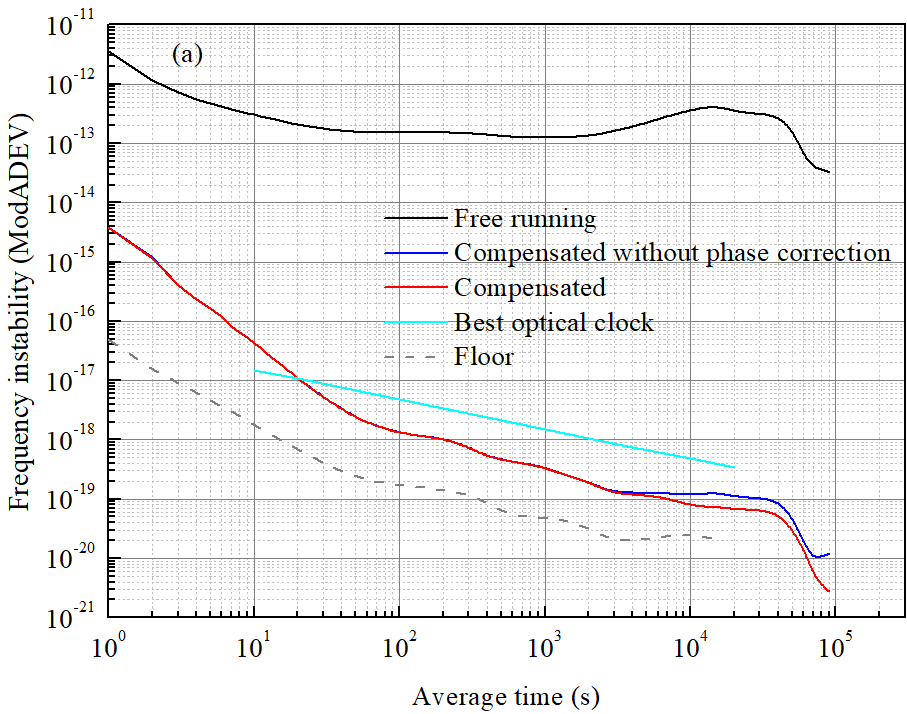}
\centering\includegraphics[width=10 cm]{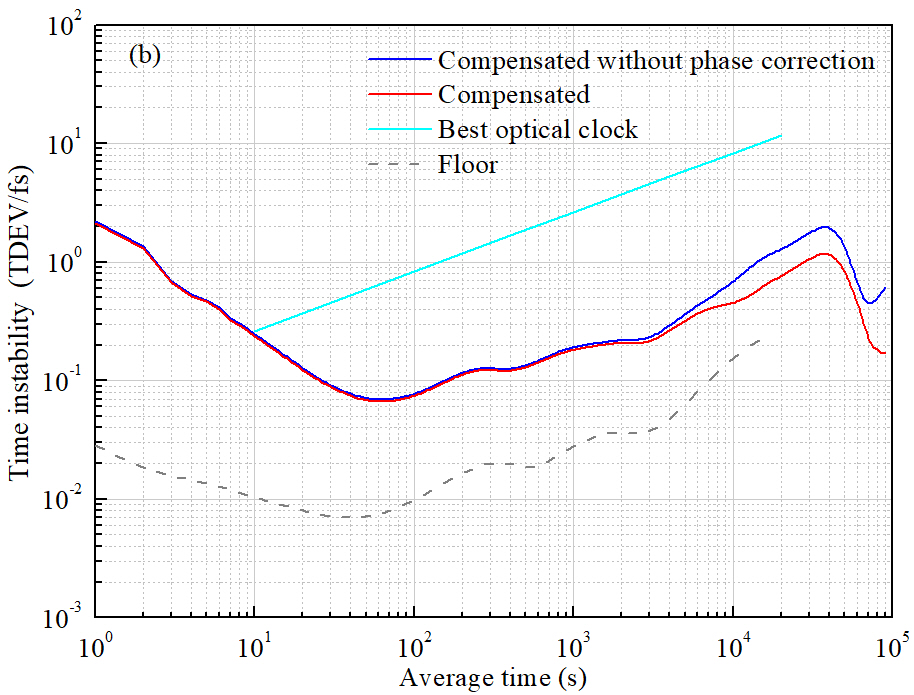}
 \caption{\textbf{(a) Fractional frequency instability (ModADEV).} Black curve: Free-running signal. Red/blue curves: Compensated signal with/without adaptive phase asymmetry Correction. Cyan curve: Best optical clock performance~\cite{aeppli2024clock}. Gray dot: System noise floor (compensated signal under simulated optical fiber attenuation). \textbf{(b) Time instability (TDEV).} Red/blue curves: Compensated signal with/without adaptive phase asymmetry correction. Cyan curve: Best optical clock performance~\cite{aeppli2024clock}. Gray dot: System noise floor.
 }
   \label{Fig:Instability}
 \end{figure}

Continuous 4-day measurements of the frequency-stabilized signal were conducted using a zero-dead-time frequency counter (K\&K FXE80, operating in $\Lambda$-mode with a 1-second gate time). The fractional frequency instability, shown in Fig.~\ref{Fig:Instability}(a), is quantified using the modified Allan deviation (ModADEV~\cite{benkler2015relation}). 1. Free-Running Regime (Black Curve): The ModADEV reveals noise levels exceeding prior field-deployed optical frequency transfer systems by orders of magnitude~\cite{predehl2012920,droste2013optical,chiodo2015cascaded,koke2019combining}; 2. Stabilized without Adaptive Phase Asymmetry Correction (Blue Curve): Achieves a ModADEV of $0.9\times10^{-20}$ at $\tau=1$ day; 3. Stabilized with Adaptive Phase Asymmetry Correction (Red Curve): Achieves a ModADEV of $2.9\times10^{-21}$ at $\tau=1$ day, two orders of magnitude lower than the best optical clock (cyan curve)~\cite{aeppli2024clock}. Fig.~\ref{Fig:Instability}(b) shows the link's time instability, quantified using time deviation (TDEV)~\cite{benkler2015relation}, with integration times ranging from 2 seconds to 1 day. The TDEV of the stabilized signal remains below 2 fs.

The achieved performance—a dissemination distance of 2,067 km and fractional frequency instability of $2.9\times10^{-21}$ at 1 day—establishes a new benchmark for field-deployed TF dissemination systems. This represents a 0.5–1.5 orders-of-magnitude improvement over prior long-haul optical frequency dissemination demonstrations~\cite{koke2019combining,predehl2012920,droste2013optical,calonico2014high,chiodo2015cascaded}, despite operating under environmental noise levels that are orders of magnitude higher.

This precision meets the stringent requirements for testing gravitational redshift predictions under Einstein’s equivalence principle~\cite{delva2017test}, a cornerstone of relativistic physics. Beyond validating fundamental theories, the adaptive architecture effectively mitigates both steady-state noise (e.g., fiber thermal drift) and transient disturbances (e.g., anthropogenic vibrations), addressing key challenges in terrestrial fiber networks.

The event-timing-based phase detection paradigm introduced here has broader applicability, including: 1. Microwave-Frequency Transfer: Enhances synchronization in radar and satellite networks~\cite{pound1946electronic}; 2. Seismic Sensing: Enables high-resolution subsurface monitoring~\cite{marra2018ultrastable,marra2022optical}; 3. Next-Generation VLBI Networks: Meets the long-term stability requirement of $1\times10^{-16}$ for very-long-baseline interferometry~\cite{clivati2017vlbi}. By addressing these demanding requirements, the proposed approach lays the foundation for significant advancements in both fundamental science and large-scale technological infrastructures, such as global astronomical observation networks and precision Earth science applications.

\section{Summary}

We have demonstrated ultra-stable time-frequency (TF) dissemination over a 2,067-km operational telecommunication network. The system operated continuously for four days without phase cycle slips and maintained full compatibility with dense wavelength-division multiplexing (DWDM) data traffic. It achieved a time instability below 2 fs for integration times from 2 seconds to 1 day and a fractional frequency instability of $2.9\times10^{-21}$ at 1 day. These results surpass prior benchmarks, even under uncompensated phase noise levels exceeding 5,000 $\mathrm{rad^{2}Hz^{-1}km^{-1}@1Hz}$ at 1 Hz—orders of magnitude higher than those in existing systems.

This achievement sets a new standard for the precision, robustness, and scalability of fiber-based TF dissemination. It enables ultra-stable TF transmission links that can withstand extreme environmental noise, such as that found in rapidly urbanizing regions where fiber segments face significant mechanical and thermal disturbances. Integrating this digitally enhanced fiber-based TF system with satellite-based intercontinental TF networks addresses critical needs for global-scale precision applications. These include redefining the SI second through optical clock networks, supporting relativistic geodesy and seismic sensing, building scalable quantum networks, accelerating dark matter detection, and advancing fundamental metrology for next-generation physics experiments.

\begin{backmatter}

\bmsection{Acknowledgment}

This work was supported by the National Key Research and Development (R\&D) Plan of China (Grant No. 2020YFA0309800), 
the Innovation Program for Quantum Science and Technology (2021ZD0300903, 2021ZD0300904, 2021ZD0300700), 
the National Natural Science Foundation of China (Grant Nos. T2125010, 12374470,
), the Key R\&D Plan of Shandong Province (Grant No. 2023CXPT105), the Chinese Academy of Sciences, Shandong provincial natural science foundation (Grant Nos. ZR2022LLZ006, ZR2022LLZ011). Q.Z. acknowledges support from the Taishan Scholar Program of Shandong Province and the XPLORER Prize from the New Cornerstone Science Foundation. F.X.C. and J.P.C. acknowledges support from the Young Expert Program of the Taishan Scholar Program of Shandong Province.

Fa-Xi Chen, Li-Bo Li and Jiu-Peng Chen contributed equally to this work.

\end{backmatter}


\bibliography{arxiv}






\end{document}